\documentclass[11pt,twoside]{article}
\usepackage{asp2014}

\aspSuppressVolSlug
\resetcounters

\begin{document}

\title{The LAMOST Data Archive and Data Release}
\author{Boliang He$^1$, Dongwei Fan$^1$, Chenzhou Cui$^1$, Shanshan Li$^1$, Changhua Li$^1$, Linying Mi$^1$
\affil{$^1$National Astronomical Observatories, Chinese Academy of Sciences (CAS), 20A Datun Road, Beijing 100012, China; \email{hebl@nao.cas.cn}}
}

\paperauthor{Boliang He}{hebl@bao.ac.cn}{}{National Astronomical Observatories, Chinese Academy of Sciences (CAS)}{Center of Information and Computing}{Beijing}{Beijing}{100012}{China}
\paperauthor{Dongwei Fan}{fandongwei@nao.cas.cn}{}{National Astronomical Observatories, Chinese Academy of Sciences (CAS)}{Center of Information and Computing}{Beijing}{Beijing}{100012}{China}
\paperauthor{Chenzhou Cui}{ccz@bao.ac.cn}{}{National Astronomical Observatories, Chinese Academy of Sciences (CAS)}{Center of Information and Computing}{Beijing}{Beijing}{100012}{China}
\paperauthor{Shanshan Li}{lich@nao.cas.cn}{}{National Astronomical Observatories, Chinese Academy of Sciences (CAS)}{Center of Information and Computing}{Beijing}{Beijing}{100012}{China}

\begin{abstract}
The Large sky Area Multi-Object Fiber Spectroscopic Telescope (LAMOST) is the largest optical telescope in China. In last four years, the LAMOST telescope has published four editions data (pilot data release, data release 1, data release 2 and data release 3). To archive and release these data (raw data, catalog, spectrum etc), we have set up a data cycle management system, including the transfer of data, archiving, backup. And through the evolution of four software versions, mature established data release system.
\end{abstract}

\section{Introduction}

The Large sky Area Multi-Object Fiber Spectroscopic Telescope (LAMOST) is the largest optical telescope in China\citep{2012RAA....12.1197C}. The size of its main mirror is 6.67 meters with 4 meters effective aperture. LAMOST is characterized by both a large field of view and large aperture.

After first light in 2008, the pilot sky survey from 2010 to 2012, and the regular sky survey is in progressing from 2012 to 2017. In the last four years, the LAMOST telescope has operated about 900 night. The raw data volume is about 18TB, and the product data contains about 4 million spectra, 5TB FITS files.

\section{Architecture}

The data cycle management system architecture shows in Figure \ref{fig1}.

\articlefigure[width=.7\textwidth]{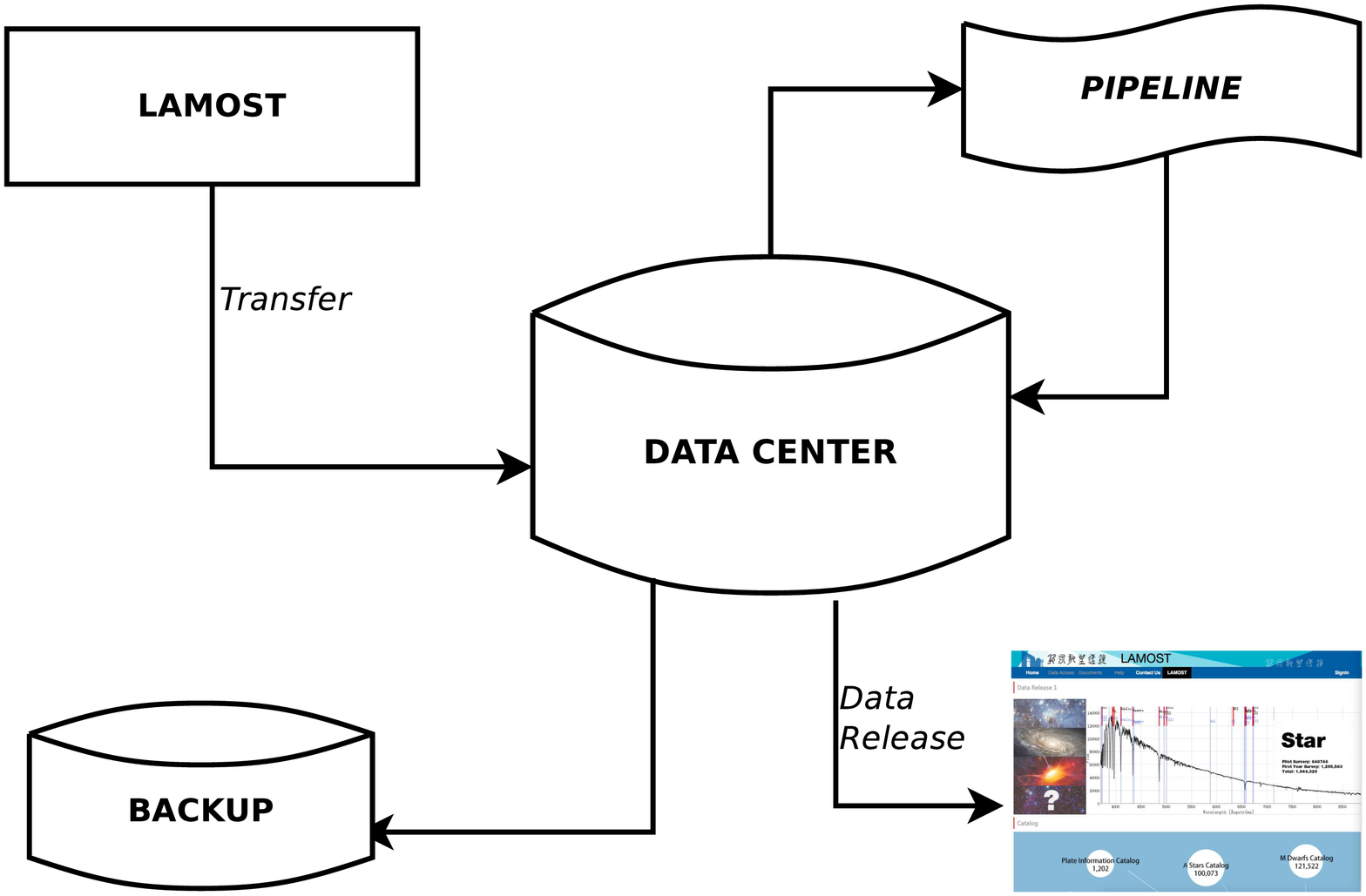}{fig1}{The data cycle management system architecture}
\section{Application}

\subsection{Data Flow}

The data flow is shown in Figure \ref{fig1}:

\begin{itemize}
\checklistitemize
\item	The site transfer raw data to China-VO Data Center. \citep{adass2014_hebl}
\item	Data Center push raw data to Pipeline Server.
\item	Pipeline generate product data (catalog and spectra FITS files) and return to Data Center.
\item	Backup raw data and product data to Backup Storage(in the third place).
\item	Push product data push to Data Release Server. \citep{adass2014_fdw}
\end{itemize}

\subsection{Data Statistic}

In the Last few years, the telescope runs about 250 days per year. Till now, LAMOST has about 900 observation nights. As show in Figure \ref{fig2}, in these 900 night, the average size of raw data is about 20GB per night.

\articlefigure[width=.7\textwidth]{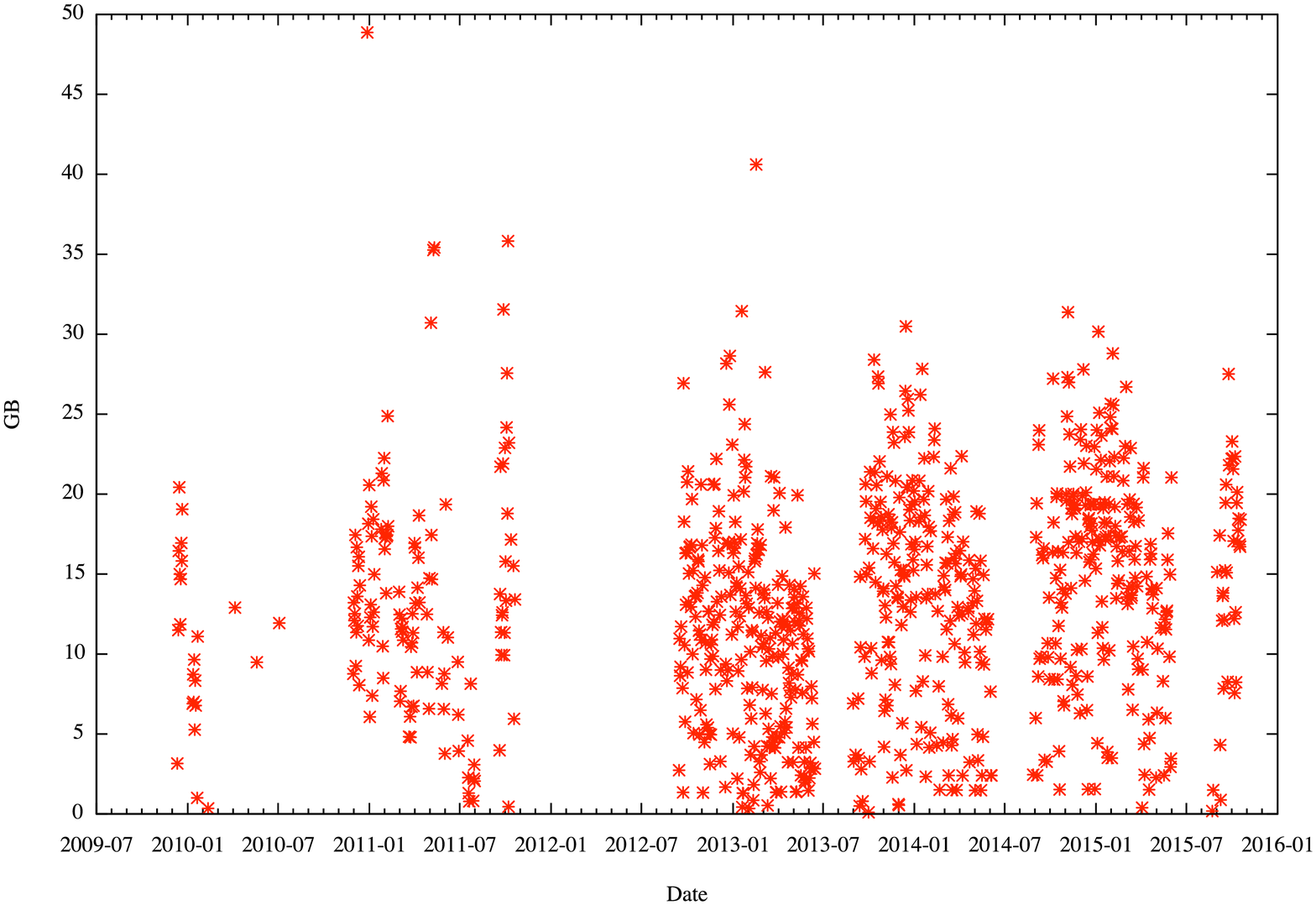}{fig2}{Raw data statistic.}

\subsection{Data Release}

After the LAMOST pipeline, the product data (catalog and spectra) will be released to users:

\begin{itemize}
\checklistitemize
\item In 2012, before the IAU General Assembly in Beijing, LAMOST Release the Pilot Data Release (PDR)\citep{2012RAA....12.1243L}
\item In 2013, LAMOST release the first Data Release (DR1)\footnote{LAMOST DR1: \url{http://dr1.lamost.org}.}, and in 2015 spring, all the DR1 data release to public.\citep{2015RAA....15.1095L}. As shown in Figure \ref{fig3}.
\item In 2014, LAMOST release the second Data Release (DR2)\footnote{LAMOST DR2: \url{http://dr2.lamost.org}.}. As shown in the left part of Figure \ref{fig4}.
\item In 2015, LAMOST release the third Data Release (DR3)\footnote{LAMOST DR3: \url{http://dr3.lamost.org}.}. As shown in the right part of Figure \ref{fig4}.
\end{itemize}

The data release software requires:

\begin{itemize}
\item Java Web Framework: Spring Framework\footnote{Spring Framework: \url{http://spring.io}}.
\item Database: PostgreSQL\footnote{PostgreSQL: \url{http://www.postgresq.org}}, pgSphere \footnote{pgSphere:  \url{http://pgsphere.projects.pgfoundry.org/}, China-VO branch: \url{https://github.com/china-vo/pgSphere}}.
\item Web Server: Nginx\footnote{Nginx: \url{https://www.nginx.org/}}.
\item User Management: CSTNET Passport\footnote{CSTNET Passport: \url{https://passport.escience.cn/}}.
\item Code Management: China-VO Code Repository Management\footnote{China-VO Code Repository Management: \url{http://code.china-vo.org}}, based on GitLab\footnote{GitLab: \url{https://www.gitlab.org/}},
\end{itemize}

\section{Conclusion}

LAMOST is an effective spectra telescope, so the data management is very challenging. Over the practical experience in past few years, we have set up a series of software and tools. In the future, We will upgrade these software and tools continually.

\acknowledgements The Guo Shou Jing Telescope (the Large Sky Area Multi-Object Fiber Spectroscopic Telescope, LAMOST) is a National Major Scientific Project built by the Chinese Academy of Sciences. Funding for the project has been provided by the National Development and Reform Commission. Data resources are supported by Chinese Astronomical Data Center(CAsDC, http://casdc.china-vo.org).

\bibliography{P051}  

\begin{thebibliography}{}
\expandafter\ifx\csname natexlab\endcsname\relax\def\natexlab#1{#1}\fi
\expandafter\ifx\csname url\endcsname\relax
  \def\url#1{\texttt{#1}}\fi
\expandafter\ifx\csname urlprefix\endcsname\relax\def\urlprefix{URL }\fi
\providecommand{\eprint}[2][]{\url{#2}}

\bibitem[{{Cui} et~al.(2012){Cui}, {Zhao}, \& et~al.}]{2012RAA....12.1197C}
{Cui}, X.-Q., {Zhao}, Y.-H., \& et~al. 2012, Research in Astronomy and
  Astrophysics, 12, 1197

\bibitem[{{Fan} et~al.(2015){Fan}, {He}, \& et~al.}]{adass2014_fdw}
{Fan}, D., {He}, B., \& et~al. 2015, in Astronomical Society of the Pacific
  Conference Series, edited by A.~R. {Taylor}, \& E.~{Rosolowsky}, vol. 495 of
  Astronomical Society of the Pacific Conference Series, 477.
  \eprint{1411.5072}

\bibitem[{{He} et~al.(2015){He}, {Cui}, {Fan}, \& et~al.}]{adass2014_hebl}
{He}, B., {Cui}, C., {Fan}, D., \& et~al. 2015, in Astronomical Society of the
  Pacific Conference Series, edited by A.~R. {Taylor}, \& E.~{Rosolowsky}, vol.
  495 of Astronomical Society of the Pacific Conference Series, 483.
  \eprint{1411.5071}

\bibitem[{{Luo} et~al.(2012){Luo}, {Zhang}, {Zhao}, \&
  et~al.}]{2012RAA....12.1243L}
{Luo}, A.-L., {Zhang}, H.-T., {Zhao}, Y.-H., \& et~al. 2012, Research in
  Astronomy and Astrophysics, 12, 1243

\bibitem[{{Luo} et~al.(2015){Luo}, {Zhao}, {Zhao}, \&
  et~al.}]{2015RAA....15.1095L}
{Luo}, A.-L., {Zhao}, Y.-H., {Zhao}, G., \& et~al. 2015, Research in Astronomy
  and Astrophysics, 15, 1095

\end{thebibliography}

\articlefigure[width=.5\textwidth]{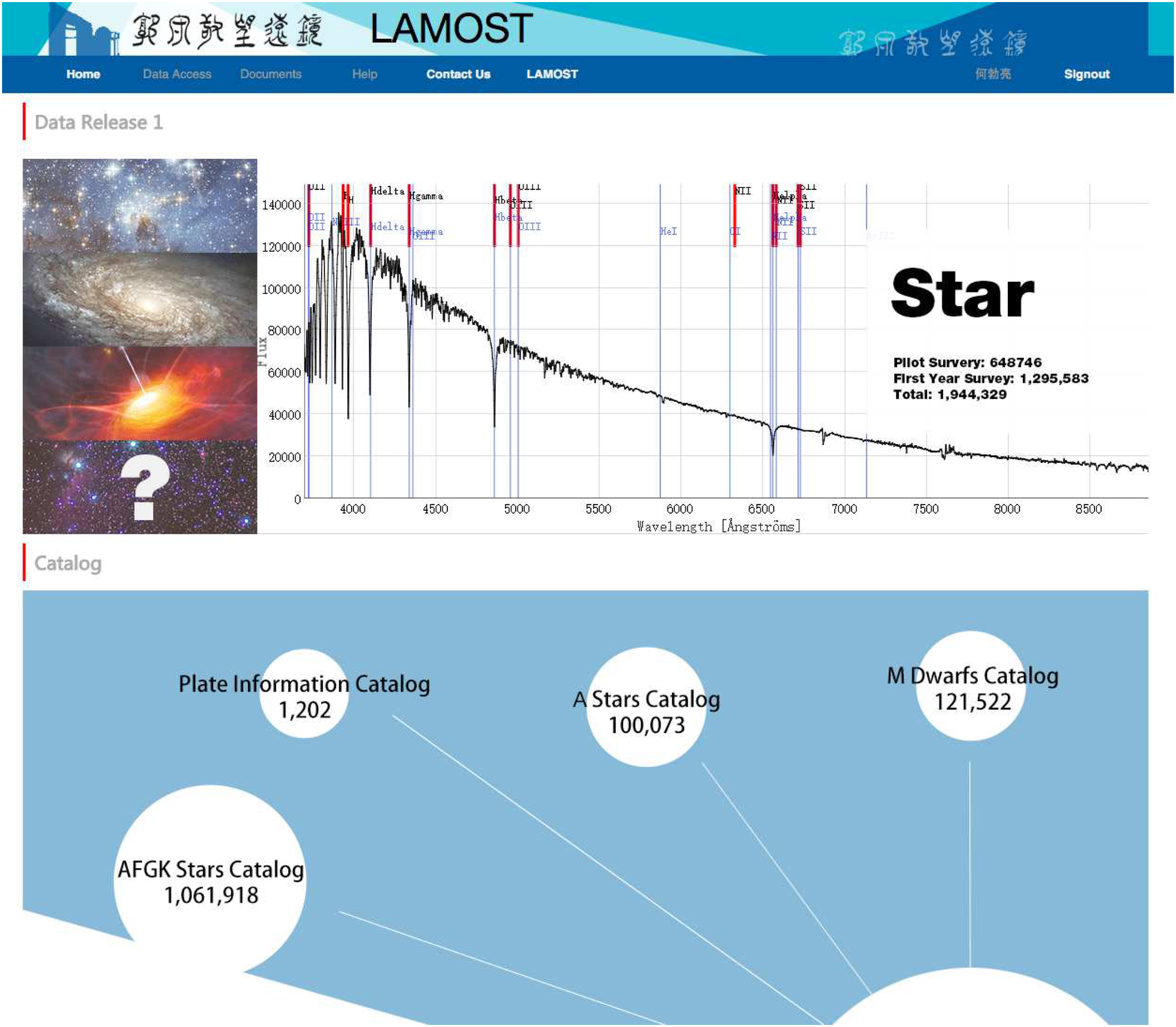}{fig3}{LAMOST DR1.}
\articlefiguretwo{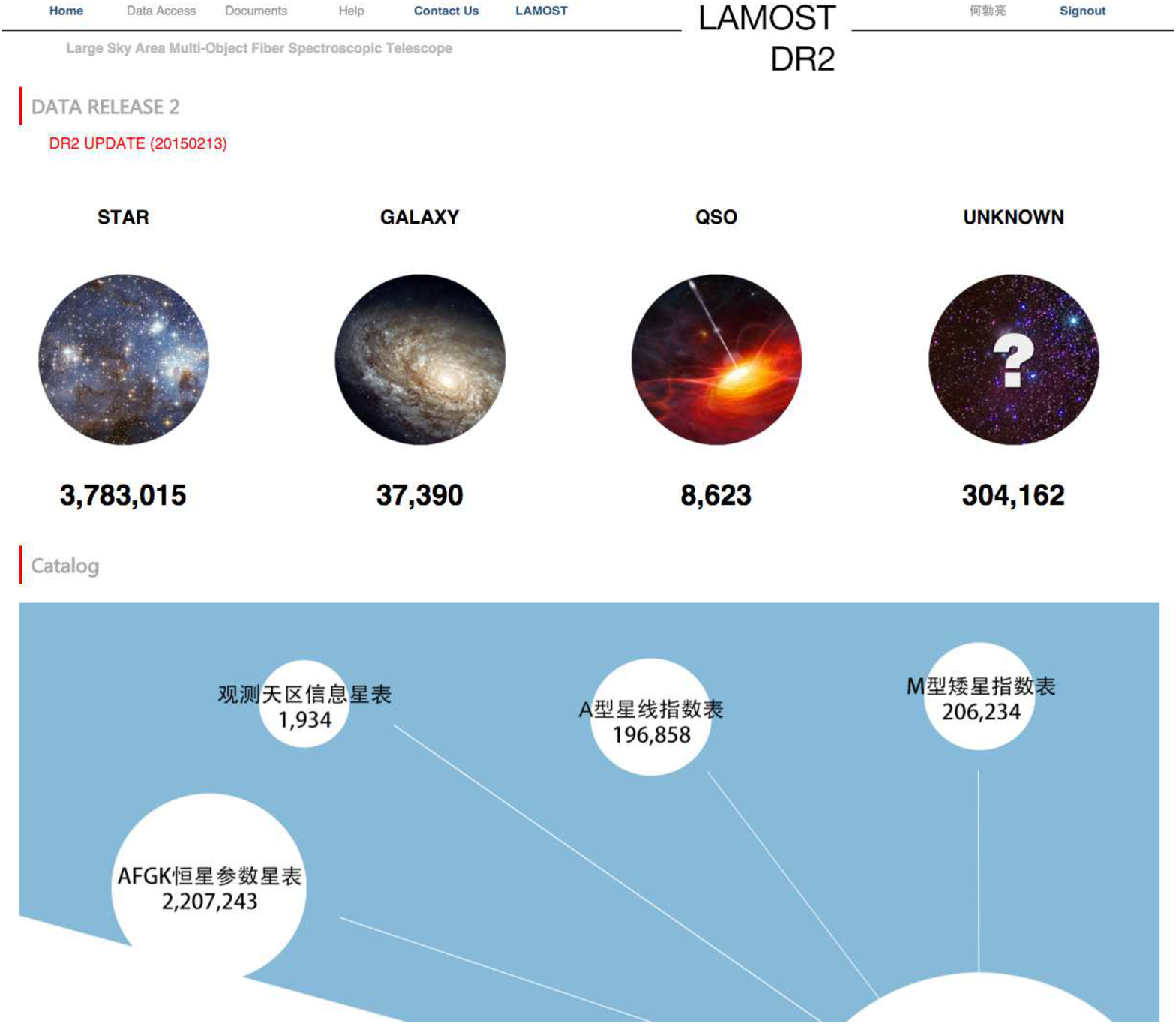}{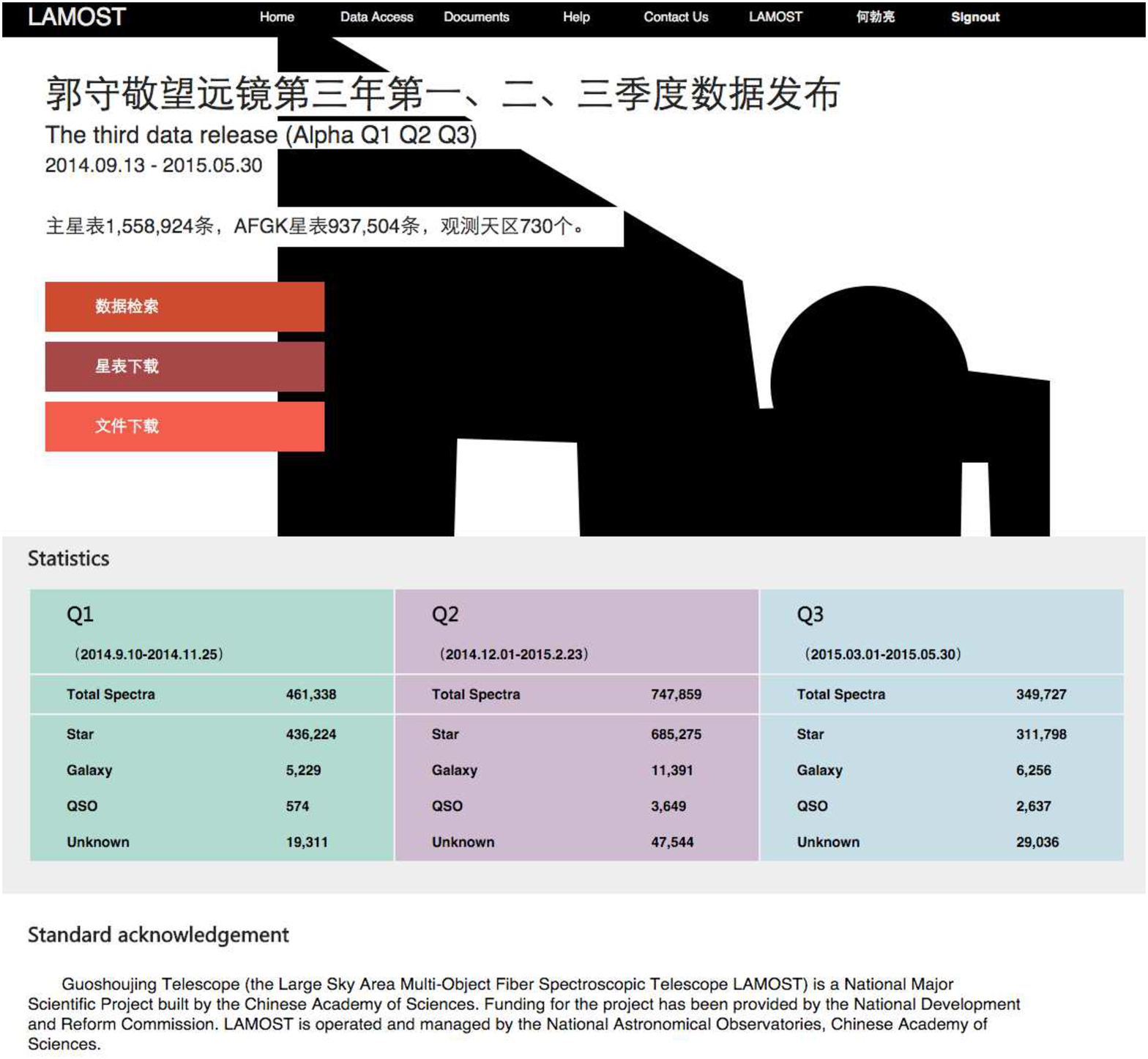}{fig4}{\emph{Left:} LAMOST DR2.  \emph{Right:} LAMOST DR3.}

\end{document}